# Nuclear charge distribution in the spontaneous fission of $^{252}$Cf

Taofeng Wang[1,2,♦], Liping Zhu[3], Liming Wang[3], Qinghua Men[3], Hongyin Han[3], Haihong Xia[3]

[1]*School of Physics and Nuclear Energy Engineering, Beihang University, Beijing 100191, China*

[2]*International Research Center for Nuclei and Particles in the Cosmos, Beijing 100191, China*

[3]*China Institute of Atomic Energy, P.O. Box 275 - 46, Beijing 102413, China*

**Abstract**

The measurement for charge distributions of fragments in $^{252}$Cf has been performed by using a unique style of detector setup consisting of a typical grid ionization chamber and a $\Delta E$–$E$ particle telescope. We found that the fragment mass dependency of the average width of the charge distribution shows a systematic decreased trend with the obvious odd-even effect. The variation of widths of charge distribution with kinetic energies shows an approximate V-shape curve due to the large number of neutron emission for the high excitation energies and cold fragmentation with low excitation energies. As for the behavior of the average nuclear charge with respect to its deviation $\Delta Z$ from the unchanged charge distribution (UCD) as a function of the mass number of primary fragments $A^*$, for asymmetric fission products $\Delta Z$ is negative value, while upon approaching mass symmetry $\Delta Z$ turns positive. Concerning the energy dependence of the most probable charge $Z_p$ for given primary mass number $A^*$, the obvious increasing tendencies for $Z_p$ with the kinetic energies are observed. The correlation between the average nuclear charge and the primary mass number is given as linear function. The proton and neutron odd-even effects with light fragment kinetic energies are derived.

## 1 Introduction

The nuclear fission reaction is a complex process involved by the nuclear dynamics in which a rearrangement of nuclear matter takes place during the descent from the saddle point to the scission. The investigations of the mass, nuclear charge and kinetic energy distributions of fragments as well as observations of correlations between these physical quantities not only can provide valuable information for

[♦]Corresponding author.
*Email address*: tfwang@buaa.edu.cn;

understanding the probability of coupling collective modes to particle excitation degrees of freedom but also contribute to explore the delicate interplay between the macroscopic aspects of bulk nuclear matter and the quantum effects of a finite nuclear system. Up to present the mass and kinetic energy distribution of fragments have been studied sufficiently for almost all known fission systems including spontaneous fission and the low energy fission reactions induced by neutrons and other light-particles. However the studies on the charge distribution of fragments, especially at the various excitation energies and various fragment mass numbers, are scarce because of the difficulty to assign the nuclear charge and the mass numbers of fission fragments simultaneously.

Isobaric charge distribution of fission fragments are represented in terms of the most probable charge $Z_p$ and the variances $\sigma$ which provide useful information regarding the dynamics of descent. The of interest characteristics is the variance of $Z_p$ and $\sigma$ dependent on the excitation energy or the total kinetic energy of fission fragments. On the other hand, more attentions have been attracted on the odd-even effects of fission products and charge polarization ($\Delta Z$) defined as the difference between the average most probable charge $Z_{pav}$ and the charge expected on the basis of Unchanged Charge Distribution ($Z_{UCD}$).

Besides the radiochemical method with the shortcoming of depending on the data of nucleus decay diagram [1], several physical approaches, such as $K$ $x$-ray intensities in coincidence with fragments [2], the energy loss of mass separated fragments in the absorber [3], $\Delta E$–$E$ particle telescope in coincidence with $K$ $x$-ray detector [4, 5], the gravity of the electron-ion pair track produced by fragments [6, 7], the prompt γ-ray in coincidence with fragments [8] and the γ-γ coincident measurement with Gammasphere [9] could be employed to determined the nuclear charge of fragments in fission reactions. It was found from checking the experimental data up to today that the charge distributions of fragments at the selected excitation energies and mass numbers in the cold region of the spontaneous fission of $^{252}$Cf have been shown by Knitter *et al.* [6]. However, probably the best way for investigating charge distributions of fragments is via the mass-spectrometer associated with the fragment residual energy detection [3, 10] or with the $\Delta E$–$E$ ionization [11], since the nuclear charge resolution of $Z/\Delta Z \sim 40$ was obtained and the variation of the charge distribution with mass numbers and kinetic energy of fragments in a wide mass region could be given by using this method [3].

A new technique employed in the present work, namely, a typical grid-ionization chamber coupled with a $\Delta E$–$E$ particle telescope [12], for determining the fragment charge distributions of spontaneous fission of $^{252}$Cf is reported. The mass of the fission fragments are determined by 2E method. The energy resolution (FWHM) of surface barrier detector for 90 MeV light fragments with mass 100 is around 1.3 MeV. In order to improve the statistics, we use 3 MeV as the kinetic energy bin. The average total kinetic energy of $^{252}$Cf(sf) <TKE> is 184.6 ± 1.3MeV. For the light fragment with kinetic energy 125 MeV, the mass resolution of the present setup is estimated as ~1.5 u on the basis of the momentum and mass conservation. The circular method of obtaining pre-neutron mass values through mean neutron multiplicities is employed in the data analysis procedure. As for nuclear charge distribution, the approach employed in Ref. [6] is that they exploited $Z$ dependence of quantity $\overline{X}$ which is the distance of the centre of gravity of the electron-ion pair track from the origin of the trace, and is in general a function of fragment energy $E$, mass $A$, and charge $Z$. The distributions of charge dependent quantity double ratio $R$ for light fragments with given mass values are shown with multi-Gaussian function fit. Concerning the present experiment, the fragment with a certain mass number and kinetic energy were selected for nuclear charge determination via the energy deposition in a ∆E-E detector. The spectrum of ∆E spectrum dependent on kinetic energies was fitted with multi-Gaussian functions which denote the yield of each charge state. The present nuclear charge resolution is better than that in Ref. [6]. The variation of the widths of the charge distributions with the fragment mass and the kinetic energy, the charge polarization as a function of the fragment mass, the dependence of the mean charge with kinetic energy and mass, as well as proton and neutron odd-even effects are discussed in the following session.

**2 Experimental procedure**

A fragment with a given mass number and the ionic charge state $q$ at a given kinetic energy was selected by the LOHENGRIN spectrograph, and the nuclear charge was determined by the residual energy of the fragment passing through an absorber [3], or by the energy deposition in a $\Delta E$ detector [11]. In contrast with this method, in the present work a detector system consisting of a grid ionization chamber (GIC) on the left side and a gas $\Delta E$ detector with a supplementary surface-barrier detector on the right side, which provide both $\Delta E$ and total kinetic energy of the fragment, was used to investigate the nuclear charge distribution of fragments of spontaneous fission of $^{252}$Cf. The diameter of the $^{252}$Cf deposition is 5mm, the fission rate is 3000 fissions/s, and the total number of coincident events recorded is about $3\times10^7$. It is indicated that the gas $\Delta E$ detector is actually a thin grid ionization chamber. The $\Delta E$-$E$ telescope utilized in this work is quite similar to that adopted in Ref. [13]. The charge distributions of the light fragments for the fixed mass number and kinetic energy were obtained by the least-squares fits for the response functions of the $\Delta E$ detector with multi-Gaussian functions representing the different elements. The results of the charge distributions for some typical fragments indicate that this detection setup has the charge distribution capability of $Z$:$\Delta Z$ > 40:1.The detailed description for experimental setup and data analysis procedure are shown in the previous publication [12]. The fits for $\Delta E_f$ using multi-Gaussian function at the conditions (a) $A^* = 86$ u, $E_L^*$=105-108 MeV and (b) $A^* = 99$ u, $E_L^* = 120$-$123$ MeV are shown in fig. 1. The width of the energy deposition spectrum of $\Delta E$ for $A^*$=125 u is very narrow, which is fitted only by three Gaussion function. The statistic of its data is not high because the yield of symmetry fission is low by several orders of magnitude. The yield of charge state for $A^*$=125 u occupy large percentage in Z=48. The structure in the distribution, FWHM of each Gaussian is around 0.8, implies that the charge resolution of $Z$:$\Delta Z$ > 40:1 has been obtained.

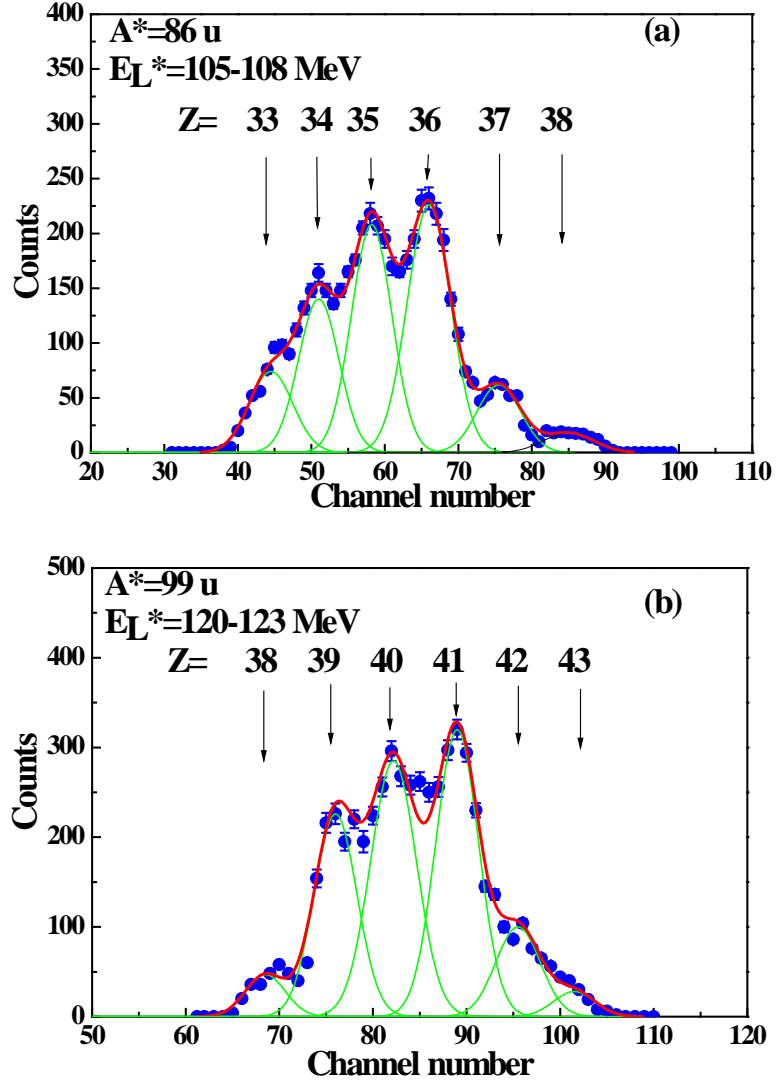

**Fig. 1.** The fit for $\Delta E_f$ using multi-Gaussian function at the conditions: (a) $A^* = 86$ u, $E_L^*$=105-108 MeV; (b) $A^* = 99$ u, $E_L^*$=120-123 MeV [12].

**3 Results and discussion**

**3.1 Isobaric yield distribution**

While the $\Delta E_f$ distributions for the selected light fragment mass $A_2^*$ and kinetic energy $E_2^*$ with an energy bin of 3.0 MeV were derived based on the analysis process as detailed in the above section, the least squares fit with multi-Gaussian distributions representing all nuclear charges was adopted. Thus the fractional independent yield (FIY) together with the charge distribution parameters, the average most probable charge $Z_{pav}$ and the dispersion $\sigma_{zav}$ of the charge distributions were obtained. Table 1, 2 and 3 show the FIY and charge distributions of light fragment $A_L^*$=101 u, 110 u and 125 u for various kinetic energies. The charge distributions depending on the different kinetic energy $E_L^*$ for the light fragments $A_2^*$=80-125 u have been obtained by using the same procedure. The uncertainties of the parameters ($Z_{pav}$, and $\sigma_{zav}$) are due to the statistics errors of the fits of the Gaussian distributions. There are a number of mass chains where it has been possible to determine the independent yields of more than one isobar. Fig. 2 shows the charge dispersion for products with $A_L^*$=101 u on 118.5 MeV kinetic energy together with a Gaussian fit for the independent yields of $^{101}$Y, $^{101}$Zr, $^{101}$Nb, $^{101}$Mo as well as $^{101}$Tc.

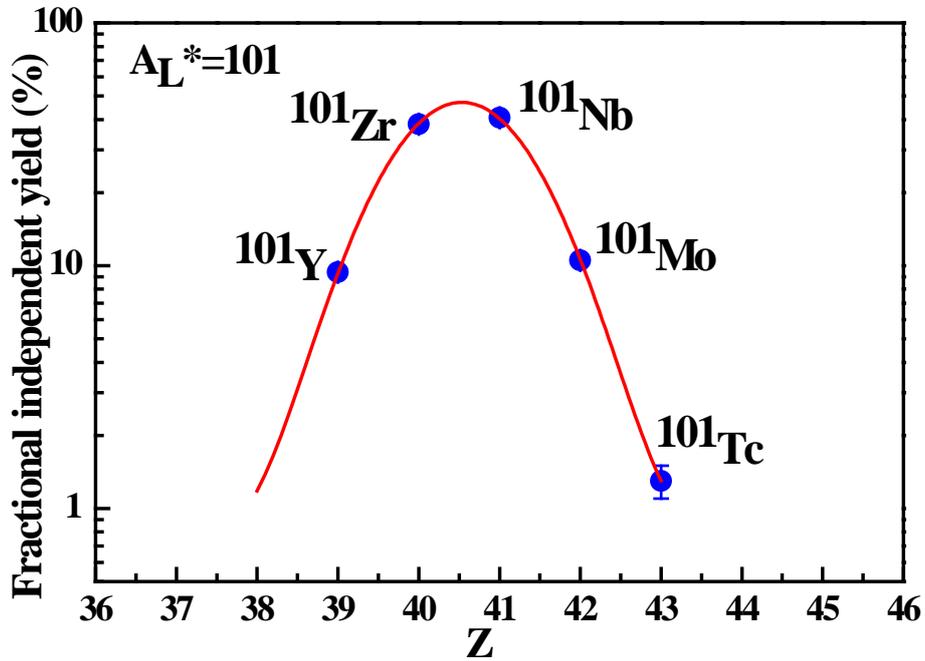

**Fig. 2.** Charge dispersion for products with $A_L^*$=101 u on 118.5 MeV kinetic energy.

Results for other mass chains exhibit a similar distribution, with the width of the distribution generally varying little with fission products mass number. Fig. 3 shows the fragment mass dependency of the average width of the charge distribution. The systematic trend is decrease when the fragments mass number increases. $d\sigma_{zav}/dA^*$ = -0.0117 has been obtained from the linear fit which is shown in fig. 3. The largest deviations occur in the mass region ~82, where nuclei have a near magic character (N=50). It is experimentally known that the average neutron multiplicities are extremely low for fragments near the shell closures, namely, the excitation energy is rather low and the total kinetic energy is rather high. On the other hand, the oscillating

Table 1. Fractional independent yields and charge distributions of light fragment $A_L^*=101$ u for various kinetic energies.

| $E_L^*$ (MeV) \ Z | 37 | 38 | 39 | 40 | 41 | 42 | 43 | $Z_p$ | $\sigma_z$ |
|---|---|---|---|---|---|---|---|---|---|
| 84 – 87 | 3.9±0.6 | 9.7±1.0 | 28.7±1.8 | 57.7±2.9 | | | | 39.40±0.36 | 0.818±0.040 |
| 87 – 90 | 1.6±0.3 | 11.1±0.8 | 29.1±1.4 | 58.3±2.2 | | | | 39.44±0.21 | 0.750±0.019 |
| 90 – 93 | 2.3±0.2 | 11.0±0.2 | 20.3±0.8 | 66.4±1.6 | | | | 39.51±0.11 | 0.779±0.013 |
| 93 – 96 | | 3.3±0.2 | 20.2±0.6 | 76.5±1.3 | | | | 39.73±0.07 | 0.512±0.005 |
| 96 – 99 | | 1.9±0.1 | 11.6±0.3 | 44.3±0.7 | 42.2±0.7 | | | 40.27±0.05 | 0.737±0.006 |
| 99–102 | | 0.57±0.05 | 8.8±0.2 | 41.6±0.6 | 49.0±0.6 | | | 40.39±0.04 | 0.669±0.004 |
| 102–105 | | 2.2±0.1 | 13.8±0.2 | 51.4±0.5 | 32.6±0.4 | | | 40.14±0.03 | 0.729±0.004 |
| 105–108 | | | 4.8±0.1 | 42.3±0.4 | 52.8±0.5 | | | 40.48±0.03 | 0.589±0.003 |
| 108–111 | | | 3.3±0.1 | 32.1±0.4 | 64.6±0.6 | | | 40.61±0.03 | 0.551±0.004 |
| 111–114 | | | | 17.6±0.3 | 74.4±0.8 | 7.9±0.2 | | 40.90±0.03 | 0.496±0.003 |
| 114–117 | | | 2.0±0.1 | 45.9±0.8 | 51.0±0.8 | 1.1±0.1 | | 40.51±0.04 | 0.559±0.004 |
| 117–120 | | | 9.4±0.5 | 38.2±1.1 | 40.7±1.1 | 10.5±0.5 | 1.3±0.2 | 40.56±0.04 | 0.849±0.008 |
| 120–123 | | | 28.7±1.6 | 41.9±2.0 | 24.5±1.5 | 3.3±0.5 | 1.6±0.3 | 40.07±0.14 | 0.897±0.021 |
| 123–126 | | | 17.3±2.2 | 36.9±3.5 | 24.0±2.7 | 13.9±2.0 | 7.9±1.4 | 40.58±0.41 | 1.16±0.10 |

$\Delta Z = -1.185 \pm 0.012$   $Z_{pav} = 40.463 \pm 0.012$   $\sigma_{zav} = 0.6009 \pm 0.0013$

Table 2. Fractional independent yields and charge distributions of light fragment $A_L^*$=110 u for various kinetic energies.

| $E_L^*$ (MeV) \ Z | 42 | 43 | 44 | 45 | 46 | 47 | $Z_p$ | $\sigma_z$ |
|---|---|---|---|---|---|---|---|---|
| 84 – 87   | 5.9±0.9   | 17.8±1.6  | 31.2±2.2  | 45.1±2.8  |           |         | 44.15±0.22 | 0.917±0.027 |
| 87 – 90   | 2.5±0.4   | 9.5±0.8   | 31.6±1.6  | 56.3±2.2  |           |         | 44.42±0.22 | 0.764±0.022 |
| 90 – 93   | 5.0±0.4   | 15.5±0.7  | 26.1±1.0  | 53.9±1.6  |           |         | 44.28±0.12 | 0.902±0.013 |
| 93 – 96   | 2.2±0.2   | 13.9±0.5  | 19.3±0.6  | 64.6±1.3  |           |         | 44.46±0.07 | 0.811±0.008 |
| 96 – 99   | 1.5±0.1   | 11.2±0.3  | 29.1±0.6  | 58.2±0.9  |           |         | 44.44±0.05 | 0.750±0.006 |
| 99–102    | 0.98±0.07 | 8.5±0.2   | 34.1±0.5  | 56.4±0.7  |           |         | 44.46±0.03 | 0.691±0.004 |
| 102–105   | 2.8±0.1   | 15.7±0.3  | 53.7±0.5  | 27.9±0.3  |           |         | 44.07±0.02 | 0.737±0.003 |
| 105–108   | 0.50±0.04 | 8.7±0.2   | 54.0±0.5  | 36.9±0.4  |           |         | 44.27±0.02 | 0.634±0.002 |
| 108–111   |           | 2.6±0.1   | 30.4±0.3  | 56.8±0.4  | 10.2±0.2  |         | 44.74±0.02 | 0.667±0.002 |
| 111–114   |           | 4.6±0.1   | 62.2±0.5  | 33.2±0.3  |           |         | 44.29±0.02 | 0.544±0.002 |
| 114–117   |           | 5.3±0.1   | 46.3±0.5  | 44.6±0.5  | 3.9±0.1   |         | 44.47±0.02 | 0.658±0.002 |
| 117–120   |           | 8.6±0.3   | 35.1±0.6  | 41.1±0.6  | 0.72±0.07 |         | 44.63±0.03 | 0.858±0.005 |
| 120–123   |           | 3.8±0.3   | 42.4±1.1  | 48.9±1.1  | 4.6±0.3   |         | 44.54±0.04 | 0.678±0.006 |
| 123–126   |           |           | 23.0±1.2  | 61.8±2.3  | 13.0±0.9  | 2.2±0.3 | 44.94±0.09 | 0.666±0.013 |
| 126–129   |           |           | 38.4±2.7  | 44.9±3.0  | 12.2±1.4  | 4.6±0.8 | 44.83±0.13 | 0.812±0.031 |

$\Delta Z = -1.526 \pm 0.008$    $Z_{pav} = 44.407 \pm 0.008$    $\sigma_{zav} = 0.6536 \pm 0.0009$

Table 3. Fractional independent yields and charge distributions of light fragment $A_L^*$=125 u for various kinetic energies.

| $E_L^*$ (MeV) \ Z | 46 | 47 | 48 | 49 | $Z_p$ | $\sigma_z$ |
|---|---|---|---|---|---|---|
| 105–108 | | 17.9±1.3 | 79.9±3.3 | 2.3±0.4 | 47.84±0.09 | 0.421±0.008 |
| 108–111 | | 11.5±0.6 | 86.1±2.3 | 2.5±0.3 | 47.91±0.05 | 0.362±0.005 |
| 111–115 | | 7.0±0.5 | 89.0±2.5 | 4.0±0.4 | 47.97±0.06 | 0.331±0.006 |
| 114–118 | | 6.1±0.8 | 88.2±3.9 | 5.8±0.8 | 48.00±0.13 | 0.344±0.039 |
| 117–120 | | 13.9±2.4 | 74.9±6.7 | 11.2±2.1 | 47.97±0.21 | 0.500±0.027 |

$\Delta Z$=0.684±0.034    $Z_{pav}$=47.927±0.034    $\sigma_{zav}$=0.3648±0.0034

nature of $\sigma_{zav}$ as a function of fragment mass [14, 15] indicates the presence of the odd-even effect. The calculation results with GEF code [16] are also given in fig. 3, for which a mostly flat trend is exhibited with fluctuations due to the known odd-even effect. Two pronounced peaks at $A^*$=82 and 126 are likely caused by shell structure.

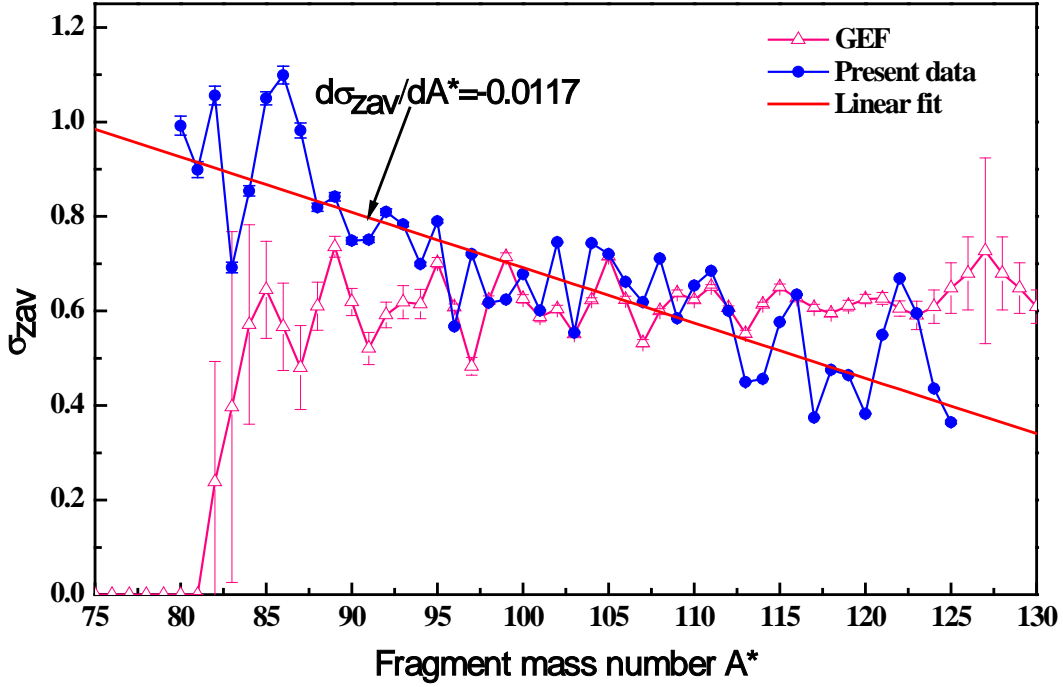

**Fig. 3.** The fragment mass dependency of average widths of the charge distributions, the solid points indicate the present experimental data, the triangles represent the calculation results from GEF code [16], the solid line is the linear function fitting.

The rms widths $\sigma_z$ of charge distributions dependent on the kinetic energy of light fragments with given masses are plotted in fig. 4. The prominent feature is that $\sigma_z$ has a decrease tendency in case the kinetic energy $E_L^*$ less than 110 MeV, correspondingly, $\sigma_z$ shows an increase trend as $E_L^*$ great than 110 MeV. In the other words, the parameters (average value and rms width of Gaussian distribution) of charge distributions vary not only on the fragment mass, but also on the excited energy of fission compound nucleus. Comparing the mass and charge distributions for the same fission reactions, one notices that asymmetric distributions are observed for both the mass and the charge yields with the peak/valley ratios for asymmetric/symmetric splits of the parent nucleus. It suggests that fragment mass and fragment charge are rather intimately linked [17]. From the broadening of the mass distributions at larger excitation energies of the compound nucleus one should also expect the global charge distributions of fragments to become wider at higher temperatures, provided that correlation between fragment mass and charge addressed above remains valid. However, as seen from fig. 4, the trends for the widths $\sigma_z$ of charge distributions are not a linear function on the kinetic energies (or excitation energies). Minimum values of $\sigma_z$ exist at around 110 MeV of $E_L^*$ which is the average kinetic energy of light fragments of $^{252}$Cf, in other words, a complex correlation was observed for the width of charge distribution and the excitation energy of fragment.

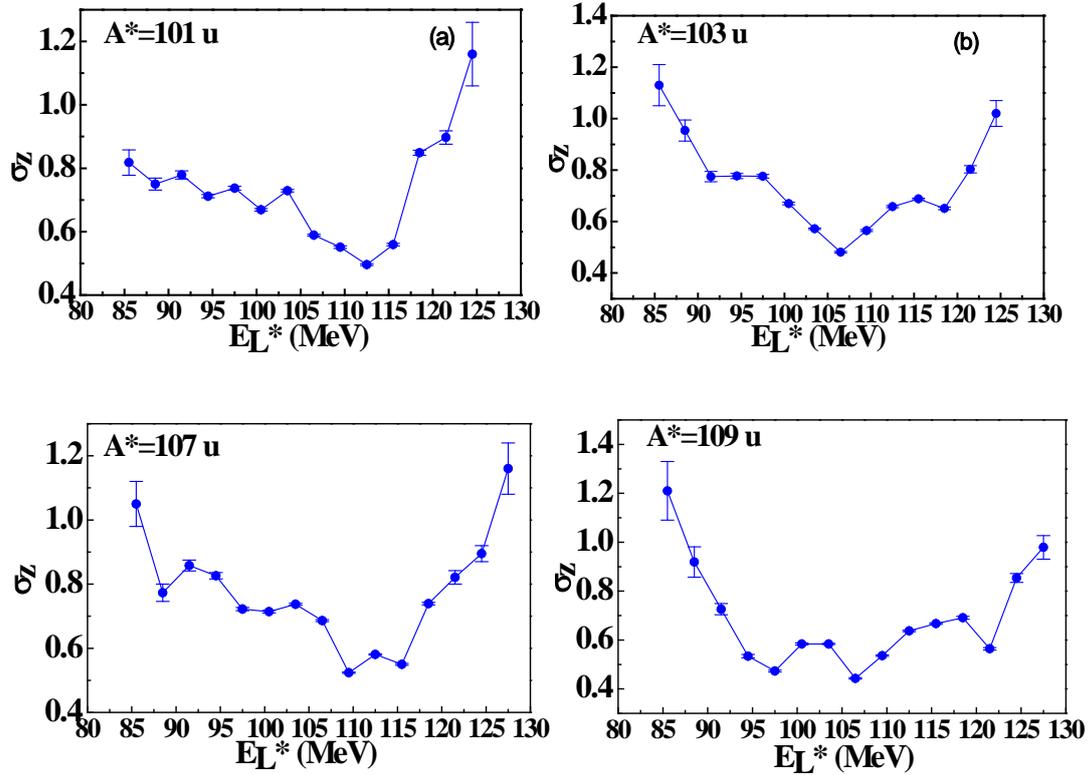

**Fig. 4.** The widths $\sigma_z$ of charge distributions depend on the kinetic energy of the light fragments $E_L^*$: (a), (b), (c) and (d) corresponding to the primary fragment masses 101, 103, 107 and 109 u, respectively.

The difference in the $\sigma_z$ variation with kinetic energy may be comprehended from the fact that for low kinetic product energies the corresponding excitation energies and, hence, the number of evaporated neutrons will be large; for large neutron number $v$ their variance $\sigma^2(v)$ is also large and, therefore, a given primary mass $A^*$ together with its intrinsic charge distribution will be spread over a wider range of product mass $A$. As a result, the smaller the kinetic energy of the fission products, the more the charge variance measured should rise compared to the variance of the primary fragments [17]. It is interesting to analyze the evolution of fragment charge distributions up to the highest feasible kinetic energies, namely, in cold fission for which fragments have such low excitation energy that no neutrons are emitted. The fragments are formed in their ground states and with ground-state deformations. The fragment elongations in fission vary over a large range and the elongation uncertainty resulting from the quantum-mechanical zero-point vibration seems to be in the order of 1-2 fm [6]. For a typical fragment split $^{109}$Tc and $^{143}$Cs in the spontaneous fission of $^{252}$Cf, the deformation energy for Tc is estimated about 6.5 MeV/fm. Fig.4 shows the $\sigma_z$ variation with kinetic energy on A*=101,103,107 and 109 u which probably represent the typical fragment splits holding large elongation uncertainty, consequently large $\sigma_z$ are produced.

During the charge equilibrium at a fixed mass value, the charge equilibrium mode *N/Z* is commonly described by a harmonic oscillator having a phonon energy $\hbar\omega$ characteristic of giant dipole resonance (a few MeV for a nucleus at scission) [18, 19]. This oscillator is coupled to the intrinsic degrees of freedom. Under these condition the variance of charge $<\sigma_z^2>$ dependent on the nuclear temperature *T* and the inertial parameter of *N/Z* mode *M*.

$$<\sigma_z^2> = \frac{1}{M\omega^2}(\frac{1}{2}\hbar\omega + \frac{\hbar\omega}{e^{\hbar\omega/T}-1}) \qquad (1)$$

Obviously, from fig. 4 a decrease can be observed with increasing (decreasing) kinetic (excitation) energy for $E_L^*$ < 110 MeV, which can be understood with the above equation. Since the nuclear temperature *T* is related to excitation energy $E_x$ by $E_x = aT^2$ with $a = \frac{1}{8}A$ from [3]. Djebara *et al*. shows a flat behavior of $<\sigma_z^2>$ for different fissioning systems [10], which is explained by the zero-point oscillation of a collective-isovector giant dipole resonance of the composite system at the exit point (the physical scission point). The dependence of $<\sigma_z^2>$ on kinetic energy should have a dynamical origin through the change of neck radius *c* with time. Nifenecker found that the asymptotic ($t \to \infty$) value of the charge variance $<\sigma_z^2>$ increases strongly with the necking velocity (d*c*/d*t*), i.e. the speed at which the neck pinches off [20, 21], where it shows $<\sigma_z^2>$ has a linear correlation with d$c$/d$t$, which account for the increase trend in highest kinetic energy ($E_L^* > 110$ MeV) in fig. 4. Brissot and Bocquet [18] have clearly exhibit that the collapse of neck is 5 times faster for $^{250}$Cf* than $^{230}$Th*. The increase of $\sigma_z$ indicates that the velocity of pinching of neck fast increase with the highest kinetic energies for cold fission.

### 3.2 Charge polarization

An important approach regarding the nuclear charge distribution in fission is to investigate the behavior of the average nuclear charge $Z_{pav}$ with respect to its deviation $\Delta Z$ from the unchanged charge distribution (UCD), as a function of the mass number of primary fragments $A^*$ or the most probable charge [22].

$$\Delta Z = (Z_{pav} - Z_{UCD})_H = (Z_{UCD} - Z_{pav})_L, \qquad (2)$$

$$Z_{UCD} = (Z_F / A_F) \times (A + \nu_A), \qquad (3)$$

where $H$ and $L$ designate the heavy and light fragments, respectively. $Z_F$ and $A_F$ are charge and mass of the fissioning system. $\nu_A$ is the number of neutrons emitted by the corresponding fission fragment. Accordingly $\nu_A$ for the heavy and light fission product mass is given as [22]

$$\nu_H = 0.531\bar{\nu} + 0.062(A_H - 143), \qquad (4)$$

$$\nu_L = 0.531\bar{\nu} + 0.062(A_L + 143 - A_F), \qquad (5)$$

where $\bar{\nu}$ is the average number of neutrons emitted during the fission process. It is taken as 3.765 for $^{252}$Cf [23].

The charge polarization on the mass number of primary fission fragments $A^*$ are shown in fig. 5, which is similar to the results of M. Djebara [10]. Djebara's work considered the thermal neutron-induced fission of $^{249}$Cf. This system is different from spontaneous fission $^{252}$Cf studied in the present work, in particular the excitation energy in thermal neutron-induced fission of $^{249}$Cf is higher (the neutron separation energy of $^{250}$Cf is 6.6 MeV). The main characteristics of $\Delta Z$ emerging from the figure are for asymmetric fission ($A_L^*<120$) $\Delta Z$ is negative value which exhibits marked fluctuations due to odd-even effects and neutron emission, while upon approaching mass symmetry ($A_L^* \geq 120$) $\Delta Z$ turns positive and does not fluctuate. The spectacular change of sign for $\Delta Z$ was not anticipated by theory. For asymmetric mass split region the fragments are polarized in charge with more than half a proton being transferred from the heavy to the light fragment. The heavy fragment with the closed proton shell $Z=50$ may play an essential role, the charge deviation $\Delta Z$ reverts sign. Even though the Minimum Potential Energy (MPE) model [24] may interpret the average more than half a proton shifted from the heavy to the light fragment as compared to an unchanged charge distribution, it could not correctly predicts the trends and fluctuations of $\Delta Z$ with mass or the change of sign of $\Delta Z$ close to mass symmetry. The size of odd-even fluctuations in $\Delta Z(A^*)$ decreases with increasing fissility $Z_F^2/A_F$ and increasing excitation energy of the fissioning nucleus ($Z_F$, $A_F$) [17]. The calculation for $\Delta Z(A^*)$ with GEF code is also shown in fig. 5, the results are generally coincident with the present data except the region near magic number ($N=50$).

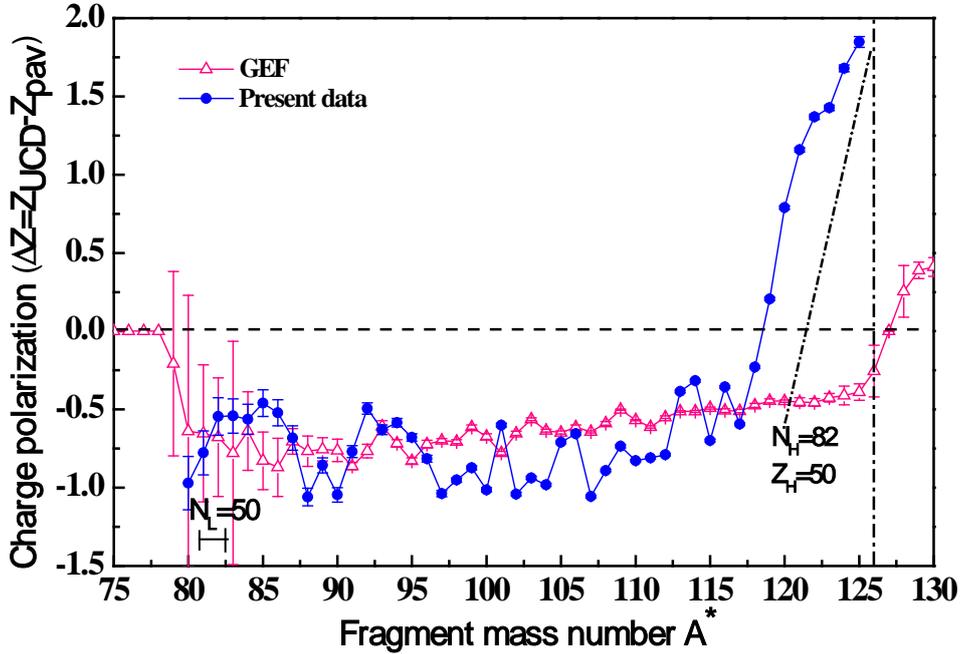

**Fig. 5.** Deviation of the average charge from the unchanged charge distribution $\Delta Z(A^*)$ as a function of the mass number of primary fission fragments $A^*$. The solid points indicate the present experimental data, the triangles represent the calculation results from GEF code [16]. The dot line indicates the neutron number and proton number for heavy fragments. The dash dot line indicates the symmetry fission.

**3.3 Energy dependence of the most probable charge**

Kinetic fragment energies dependence of the most probable charge $Z_p$ are shown in fig. 6 where the primary mass number of light fragment are given. The obvious increasing tendencies for $Z_p$ with the kinetic energies are emerged from the figure which is measured for the first time. It can be seen that $Z_p$ of light fragment mass chains vary with energy in similar ways. It is likely that the nuclear charge distribution between the primary fragments is controlled by a process that is unaltered by additional excitation in compound nucleus; hence, the resultant change in $Z_p$ with kinetic energy (or excitation energy) is due only to increased neutron emission at higher excitations [25]. This similar variation in $Z_p$ with energy adds support to the assumption of an identical charge dispersion in the mass chains. The energy ($E_L^*>110$ MeV) is well above average and, in fact, one is approaching cold fission, where all of the available reaction energy $Q(A, Z)$ is converted into fragment kinetic energy. Upon coming close to cold fission, it is found that in most cases the charge number $Z(A)$ maximizing the yield $Y(Z|A)$ for given mass number A coincides with the charge maximizing reaction energy $Q(Z|A)$. The shift of $Z_p$ of charge distribution in a given isobaric chain is connected only with prompt neutron emission, which shifts fragment nucleus closer to the valley of β-stability. Secondly, it is related with a decrease of the charge polarization in primary fission fragments, which leads to a decrease of charge density of light fragments and an increase of charge density of heavy fragments [26]. As the excitation energy of the compound nucleus increases the primary heavy fragments therefore are approaching the line of *β*-stability and primary light fragments are moving off it. This effect reduces the most probable charge of light fission products and raises it for heavy products.

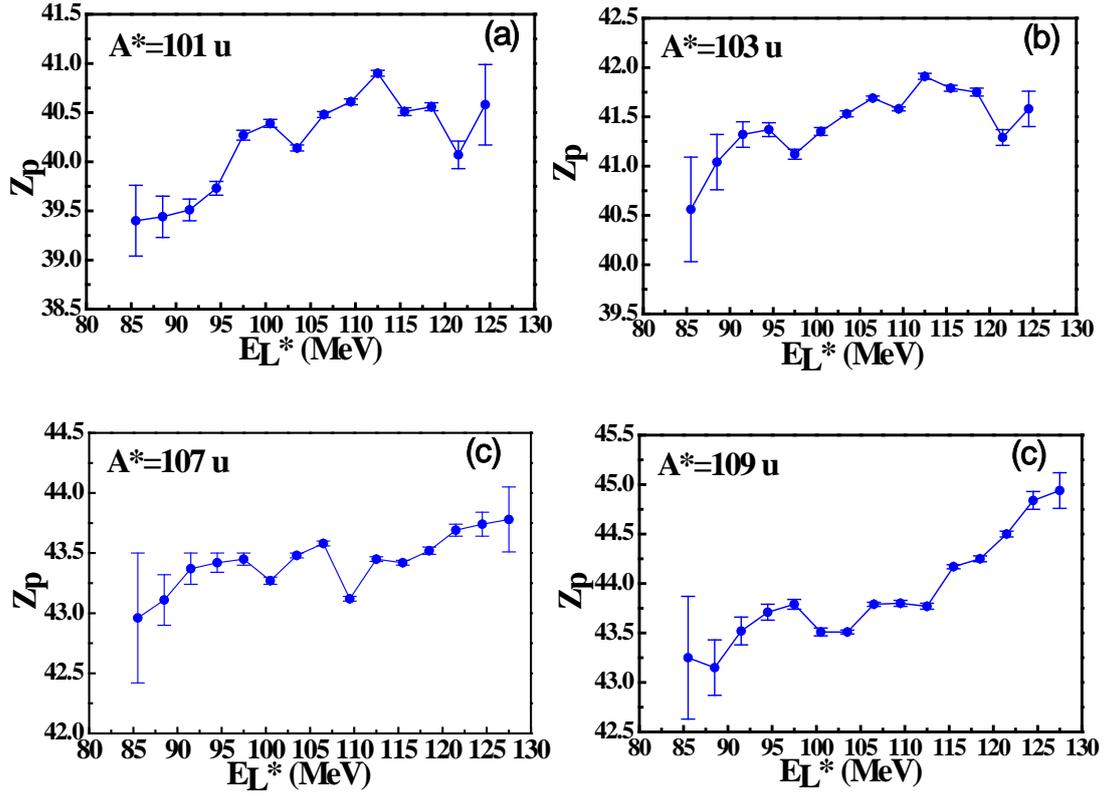

**Fig. 6.** Kinetic fragment energies dependence of the most probable charge $Z_p$ for $^{252}$Cf: (a), (b), (c) and (d) corresponding to the primary fragment masses 101, 103, 107 and 109 u, respectively.

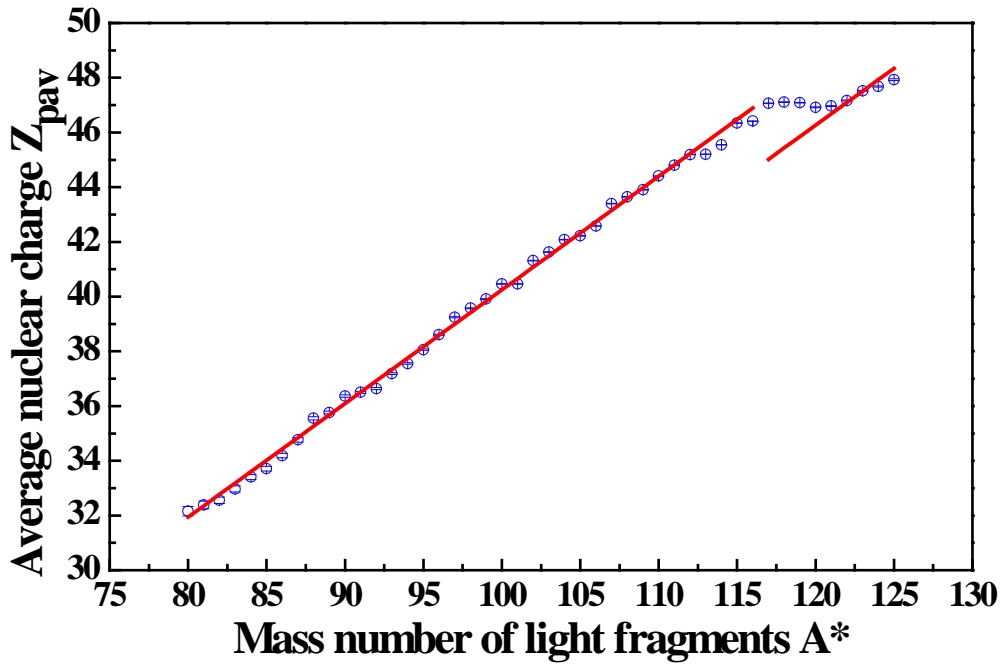

**Fig. 7.** The average nuclear charge $Z_{pav}$ as a function of mass number of light fragments for $^{252}$Cf. The solid curve is the results of calculations for Waldo's $Z_{pav}(A^*)$ formula.

The average charge of isobaric chains with the mass number of light fission products are obtained, as shown in fig. 7. The results of calculations for Waldo's $Z_{pav}(A^*)$ formula [27] are also exhibited in the figure.

$$Z_{pav}(A^*) = 0.4153 \cdot A - 1.19 + 0.167 \cdot (236 - 92 \cdot A_F / C_F),$$

A<116  (6)

$$Z_{pav}(A^*) = 0.4153 \cdot A - 3.43 + 0.243 \cdot (236 - 92 \cdot A_F / C_F),$$

A>116  (7)

It can be seen from this figure that the present data are in a perfect agreement with the results calculated with Waldo's formula [27].

### 3.4 Odd-even effect

A correlation between the odd-even effect and fissility is apparent, the proton odd-even effect [10] can be expressed by a function of the Coulomb parameter $z = Z_F^2 / A_F^{1/3}$ for $\delta_p[\%] = ae^{-bz}$ where $a = e^{17.5}$ and $b = 0.01049$. The proton odd-even effect of spontaneous fission nucleus $^{252}$Cf is not conspicuous to compare to that of $^{230}$Th and $^{233}$U [10]. It is worthwhile to investigate the odd-even fluctuations of fragment average charge yield as a function of fragment kinetic (or excitation) energy. The odd-even staggering of average charge yield in the light fragment group increases with the light fragment kinetic energy $E_L^*$, as is shown in fig. 8. At high kinetic energies a large percentage of even atomic numbers is present than at low energies. In the other words, fragments with even charge numbers $Z$ will on average exhibit higher kinetic energies than those with odd $Z$. The average charge yields of $Z_L = 46$ (or 47) occupying large percentage is palpably visible. One may speculate whether this feature is influenced by the closed $Z_H = 50$ proton shell of heavy fragments. The proton odd-even effect $\delta_p$ in present measurement and that of Ref. [28] are shown in Table 1.

Table 1. The proton odd-even effects of $^{252}$Cf in present measurement and Ref. [28].

| $\delta_p$ | Present results (%) | Ref. [28] (%) |
|---|---|---|
| $\dfrac{\overline{\sum Y_e} - \overline{\sum Y_o}}{\overline{\sum Y_e} + \overline{\sum Y_o}}\%$ | 8.45±0.57 | 12.0±2.0 |

The average yields of light fragment group as a function of neutron number $N$ for different kinetic energies are shown in fig. 9. For the near-barrier fission reaction, the nucleus is expected to be perfectly paired at the saddle point since not enough energy is available at that point to break pairs. Neutron odd-even effect has received less attention because any primary effects are masked by neutron evaporation. The neutron odd-even effect is rather small, the rise is spectacular for large fragment kinetic energy. At high kinetic energy the fragment excitation energies exhibiting the emission of neutron and gamma will be low. Hence, the energy dependence neutron odd-even effect points to a huge primary neutron effect for very asymmetry fission. Fission with very asymmetric mass splits carries the signature of cold deformed fission in which fragment kinetic energies are high, and odd-even effects are pronounced. To some extent, the limited mass resolution of experimental setup impacts the odd-even effects, especially for the very asymmetric fission which has low yield by several order of magnitude.

In fig. 9 the low kinetic energy corresponding to high excitation energy, the neutron odd-even effect has a distinct fluctuation due to the prompt neutron evaporation. Pronounced peaks exist in the region of neutron number from 67 to 75 for intermediate and high kinetic energy. The possible reason for the peaks is the $N_H =$ 82 neutron shell effect in heavy fragments production. Regarding to the high kinetic energy case, one distinct peak locates at $N$=57 which also appear at the same place for intermediate energy case. None of the pronounced neutron odd-even effects for light fragments in very asymmetric fission are observed due to the low yield in this region. If the mass resolution is improved by using the E-V or 2V method in the fission TOF spectrometer, the odd-even effects will be more clear and visible.

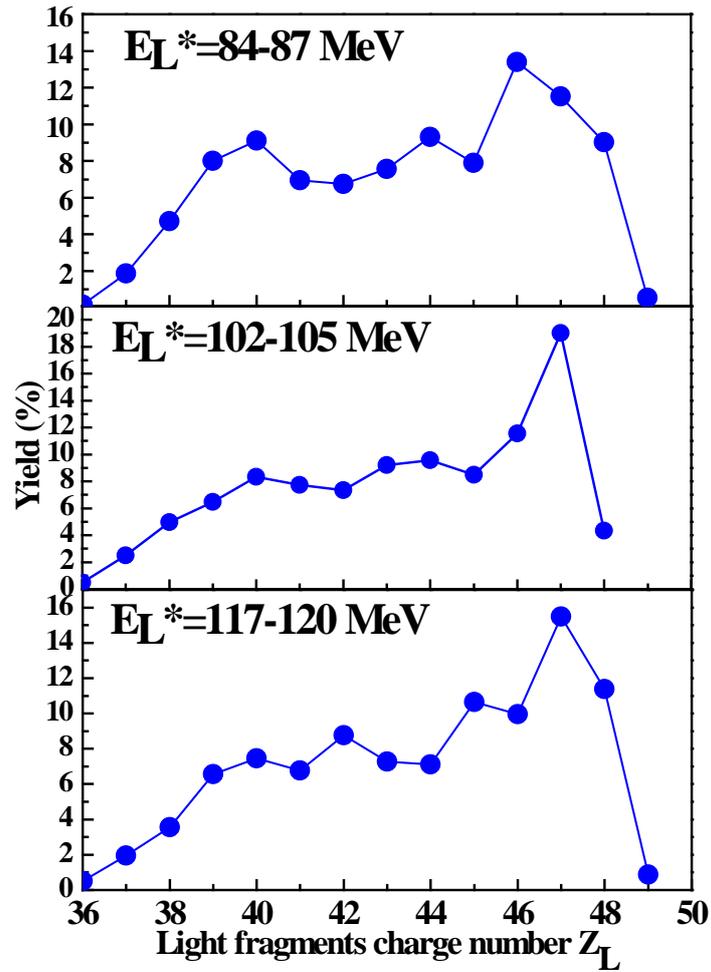

**Fig. 8.** The yield of the light fragment group on charge number $Z_L$ with the light fragment kinetic energy $E_L^*$.

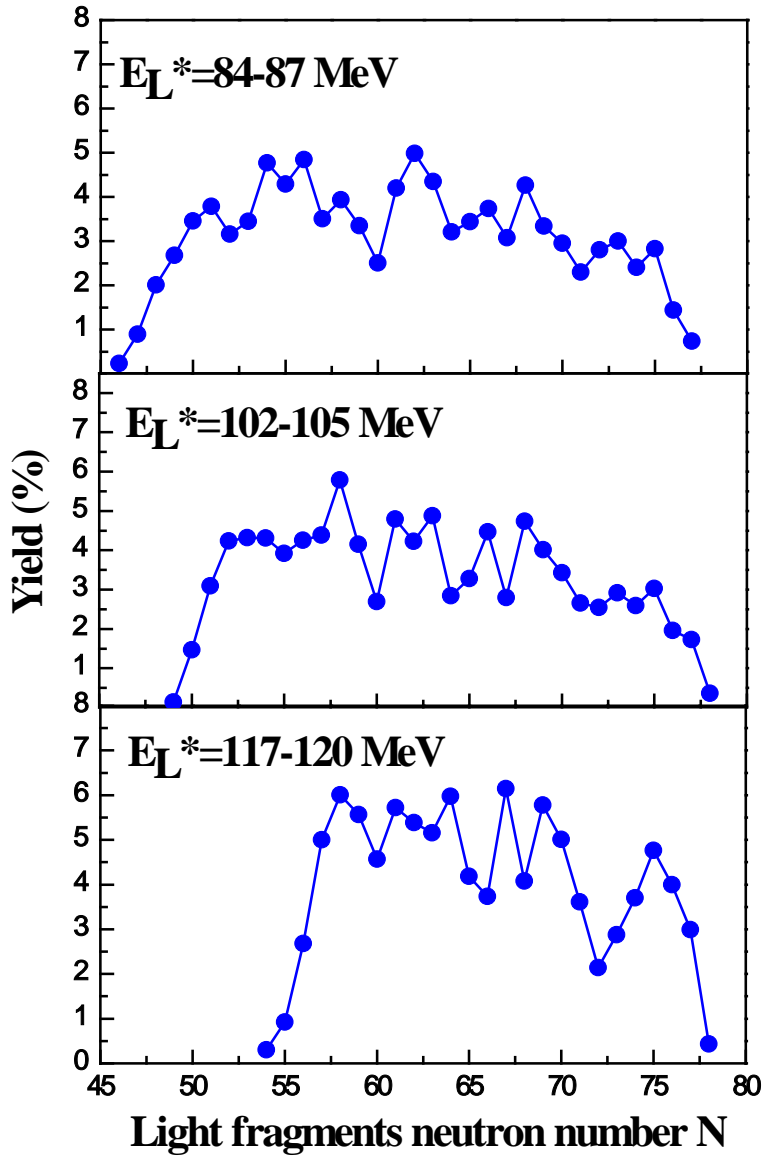

**Fig. 9.** The yield of the light fragment group on neutron number *N* with the light fragment kinetic energy $E_L^*$.

## 4. Summary

The measurement for charge distributions of fragments in spontaneous fission $^{252}$Cf has been performed by using a unique style of detector setup consisting of a typical grid ionization chamber and a *ΔE–E* particle telescope，in which a thin grid ionization chamber served as the *ΔE*-section and the E-section was a surface barrier detector. The typical physical quantities of fragments, such as mass number and kinetic energies as well as the deposition in the gas *ΔE* detector and E detector were derived from the coincident measurement data. The charge distributions of the light fragments for the fixed mass number and kinetic energy were obtained by the least-squares fits for the response functions of the $\Delta E$ detector with multi-Gaussian functions representing the different elements. The results of the charge distributions for some typical fragments are shown in this article which indicates that this detection setup has the charge distribution capability of $Z:\Delta Z > 40:1$. The experimental method developed in this work for determining the charge distributions of fragments is expected to be employed in the neutron induced fissions of $^{232}$Th and $^{238}$U or other low energy fission reactions.

As a result of the study the fragment mass dependency of the average width of the charge distribution shows a systematic decreased trend with the obvious odd-even effect. As for the variation of widths of charge distribution with light fragment products kinetic energies, the approximate V-shape curves have been observed due to the large number of neutron emission for the low kinetic energies (or the high excitation energies), and the approaching cold fission for high kinetic energies ($E_L^*>107$ MeV). In the case of the behavior of the average most probable nuclear charge $Z_{pav}$ with respect to its deviation $\Delta Z$ from the unchanged charge distribution (UCD) as a function of the mass number of primary fragments $A^*$, for asymmetric fission products ($A_L^*<120$) $\Delta Z$ is negative value which exhibits marked fluctuations due to odd-even effects, while upon approaching mass symmetry ($A_L^*\geqslant 120$) $\Delta Z$ turns positive and does not fluctuate. It is likely that for asymmetric mass split region the fragments are polarized in charge with more than half a proton being transferred from the heavy to the light fragment. The heavy fragments with the closed proton shell $Z=50$ may play an essential role, the charge deviation $\Delta Z$ reverts sign. Concerning the energy dependence of the most probable charge $Z_p$ for given primary mass number $A^*$, the obvious increasing tendencies for $Z_p$ with the kinetic energies $E_L^*$ are observed. The reason causes this phenomenon is due only to increased neutron emission at higher excitations. The correlation between the average nuclear charge and the primary mass number is given as linear function, which is in a perfect agreement with the calculation results from Waldo's formula. The proton and neutron odd-even effects with light fragment kinetic energies are derived. The odd-even staggering of charge yield in the light fragment group increases with the light fragment kinetic energy.

## 5. Acknowledgements

The authors would like to express their sincere thanks to Prof. Cheng Yu for designing and making the GIC as well as provide very nice advices for testing the whole detection setup. This work has been supported by the National Natural Science Foundation of China (No. 10175091, No.11235002 and No. 11305007) and by the Fundamental Research Funds for the Central Universities.